\documentclass{nature}
\usepackage{csquotes}
\usepackage{graphicx}
\makeatletter
\let\saved@includegraphics\includegraphics
\AtBeginDocument{\let\includegraphics\saved@includegraphics}
\renewenvironment*{figure}{\@float{figure}}{\end@float}
\makeatother
\usepackage{graphics}
\newcommand{\beginsupplement}{%
	\setcounter{table}{0}
	\renewcommand{\thetable}{S\arabic{table}}%
	\setcounter{figure}{0}
	\renewcommand{\thefigure}{S\arabic{figure}}%
}

\title{Magnetic-field-orientation-dependent triplet supercurrents in Josephson junctions with symmetric and asymmetric exchange-spring interfaces}

\author{Ekta Bhatia$^{1,2,a}$, J. M. Devine-Stoneman$^{2}$, S. Komori$^{2}$, A. Srivastava$^{2}$, N. A. Stelmashenko$^{2}$, Z. H. Barber$^2$, K. Senapati$^{1,b)}$ \& J. W. A. Robinson$^{2,c)}$}

\begin{document}

	\maketitle
	
	\begin{affiliations}
		\item School of Physical Sciences, National Institute of Science Education and Research (NISER), HBNI, Bhubaneswar, Odisha, 752050, India
		\item Department of Materials Science \& Metallurgy, University of Cambridge, 27 Charles Babbage Road, CB3 0FS, United Kingdom
	\end{affiliations}

\begin{abstract}
Josephson junctions with \textit{s}-wave superconductors (S) and multiple ferromagnetic (F) layers carry spin-triplet  supercurrents in the presence of magnetically inhomogeneous spin-mixer interfaces.
Here, we report magnetic Josephson junctions with exchange-spring and double exchange-spring (\enquote{spin-mixer}) Co/Py interfaces. At the Co/Py interface, exchange coupling is strong with respect to the magnetic anisotropy of Py and so, depending on the direction and magnitude of an external magnetic field, a non-collinear magnetic structure forms in Py creating the necessary magnetic inhomogeneity for singlet-to-triplet pair conversion. We detect supercurrents through Py with a thickness exceeding 10 nm, which is much larger than the singlet pair coherence length, suggesting the propagation of magnetically tuneable triplet supercurrents.
\end{abstract}

\newpage	
Spin-singlet Cooper pairs rapidly decay within a few nanometers in a ferromagnet due to the depairing effect of a magnetic exchange field\cite{1}. Over a decade ago spin-singlet-to-spin-triplet conversion was predicted at magnetically inhomogeneous superconductor(S)/ferromagnet(F) interfaces\cite{2}. If the magnetization at the interface is parallel to the magnetization of the F layer, spin-zero triplet pairs form through a spin-mixing process\cite{3,4,5}. The decay length of spin-zero triplets match spin-singlets in Fs but spin-triplets are rotationally variant and hence, if the interface magnetization is misaligned with respect to the F, spin-polarized triplet pairs form through a spin-rotation process and these are long-ranged in the F\cite{3,4,5}. Spin-triplet pair correlations are odd in frequency and even in orbital angular momentum, and, therefore are insensitive to impurity scattering\cite{3,4,5}. 

Triplet supercurrents carry a net spin in addition to charge and so their creation is key to superconducting spintronics\cite{3}. The first experimental demonstration of spin-triplet supercurrents was reported in the half-metallic ferromagnet CrO$_{2}$ by Keizer \textit{et al.}\cite{8} in 2006. Later, several other groups reported triplet supercurrents in all-metallic Josephson junctions with transition metal Fs with different types of spin-mixer interfaces\cite{8,10,11,14,15,16,17,18}. Khaire \textit{et al.}\cite{10} used PdNi as a spin-mixer, while Robinson \textit{et al.}\cite{11} used Ho. More recently, triplet pairing has also been demonstrated in spin-pumping experiments which involve Pt/Nb/Py multilayer\cite{46,47}.

All previous triplet junction experiments are broadly based on SF$'$FF$'$S Josephson junctions\cite{20} in which the magnetizations of F$'$ and F are misaligned with respect to each other. In a SF$'$FF$'$S junction, one F$'$/F interface converts singlet pairs to triplet pairs while the second F/F$'$ interface performs the opposite process, establishing Josephson coupling between the S layers. In an asymmetric SF$'$FS junction, the first F$'$/F interface converts singlet pairs to triplet pairs, but triplet pairs decay in S over a singlet coherence length. Therefore, the mechanism for Josephson coupling in asymmetric junctions is unclear although theoretical reports\cite{21,22} propose that Josephson coupling is established through a double Andreev reflection process. 

Control and tuning of triplet supercurrent generation is an important goal for applications. Banerjee \textit{et al.}\cite{26} demonstrated reversible triplet supercurrents in Nb/Py/Cu/Co/Cu/Py/Nb junctions while Iovan \textit{et al.}\cite{17} used Nb/CuNi/Cu/CuNi/Nb spin-valve junctions to control supercurrents. In these reports, evidence for spin-triplet generation appeared during the magnetization reversal process while sweeping an external magnetic field. Other groups have reported similar control using S/F/F' spin-valves\cite{27,29,31,32,48,49}. More recently, Martinez \textit{et al.}\cite{33} demonstrated on/off control of spin-triplet supercurrents in multi-ferromagnet S/F/S Josephson junctions.

To control triplet supercurrents in S/F/S junctions, it is advantageous to be able to tune the degree of magnetic inhomogeneity present at S/F interfaces. Magnetic exchange-spring (XS) interfaces have the potential to offer such control\cite{23} and consist of neighboring magnetically hard and soft ferromagnetic materials in which a non-collinear magnetic structure is formed in the magnetically soft layer with small ($<$50 mT) external magnetic fields\cite{24}. The non-collinear magnetic structures are tuneable as a function of the direction of applied magnetic field as recently demonstrated in SmFe/Py\cite{39} and SmCo/Py\cite{40}, albeit using large magnetic fields ($\geq$300 mT).

Here, we report supercurrent-control in Josephson junctions with symmetric and asymmetric Co/Py XS interfaces in which long-range Josephson coupling is observed over Py thickness of 11 nm, which is an order of magnitude larger than the singlet coherence length in Py ($\sim1.4$ nm)\cite{50,25}. The maximum Josephson critical current is also dependent on magnetic field orientation.

\section*{Results}
 \textbf{Magnetic characterization.}
Figure 1 shows a schematic illustration of a nanopillar Josephson junction along with the in-plane magnetization versus magnetic field \textit{M(H)} hysteresis loop of the unpatterned Nb/Co(2 nm)/Py(11 nm)/Nb multilayer at 10 K. The Co is magnetically hard with respect to Py\cite{35,36}. For \textit{H} antiparallel to the Co moment [Fig. 1 (c)], a non-collinear magnetic structure forms in Py due to interfacial magnetic coupling between Co and Py. Fig. 1(a) shows the major and minor \textit{M}(\textit{H}) loops. For the minor \textit{M}(\textit{H}) loop, \textit{M} is varied from positive saturation to a field lower than the negative saturation and is then reversed back to positive saturation field. In Fig. 1(b), we show d\textit{M}/d\textit{H} of this sample, in which a plateau in d\textit{M}/d\textit{H} is observed, indicating a misalignment between the magnetic moments of Co and Py in the reversal region of the \textit{M}(\textit{H}) curve. Moreover, the minor loops are reversible up to -2 mT whereas irreversibility initiates on reversing from -3.5 mT as shown in Fig. 1(b). This shows that the non-collinear magnetic structure in Py is tuneable with magnetic fields of only a few mT. 

 \textbf{Double exchange-spring junction characteristics.} In Figure 2(a), we have plotted the in-plane \textit{I}$_{\mathrm c}$(\textit{H}) behavior of a Nb/Py(7 nm)/Co(2 nm)/Py(7 nm)/Nb junction at 1.6 K, with \textit{H} applied in-plane. \textit{I}$_{\mathrm c}$ is defined as the current at which d\textit{V}/d\textit{I} reaches a maximum. The Fraunhofer modulations of \textit{I}$_{\mathrm c}$(\textit{H}) is a characteristic feature of a Josephson junction. The \textit{I}$_{\mathrm c}$(\textit{H}) behavior confirms that this structure acts as a magnetic Josephson junction. 
 
 Figure 2(b) (inset) shows a schematic illustration of the measurement setup and \textit{I}$_{\mathrm c}$($\theta$) for the Nb/Py(7 nm)/Co(2 nm)/Py(7 nm)/Nb Josephson junction at 1.6 K and 20 mT, where $\theta$ is defined as the in-plane angle of applied magnetic field with respect to the length of the track (x-axis). For different values of H, we can tune $\theta$. For the 20 mT measurement, an initial magnetic field of 400 mT was applied to saturate Co and Py and then the field was reduced to 20 mT along the length of the track. The junction was rotated in-the-plane of the substrate in this constant applied magnetic field, as shown in Fig. 2(b) where sample rotation is equivalent to the rotation of the direction of applied magnetic field. Henceforth, we will refer to sample rotation as field rotation throughout the text.  As $\theta$ changes, the relative angle between the Co and Py layer moments changes, and hence the magnitude of the triplet supercurrent varies. The different angles of rotation correspond to different non-collinear magnetic structures in Co/Py. When singlet supercurrent passes through the magnetic structure having a particular twist, corresponding to a specific $\theta$, a fraction of the singlet supercurrent converts to a triplet supercurrent. We observe a minimum \textit{I}$_{\mathrm c}$ at $\sim$0$^{\circ}$  and $\sim$180$^{\circ}$ whereas maximum \textit{I}$_{\mathrm c}$ corresponds to $\sim$90$^{\circ}$  and $\sim$270$^{\circ}$  of applied magnetic field as shown in Fig. 2(b). The twisting of the non-collinear magnetic structure may be a maximum at certain angles, giving maximum triplet supercurrent, as reported earlier\cite{39}. We note that \textit{I}$_{\mathrm c}$($\theta$) does not drop to zero, which implies residual magnetic non-collinearity between Co and Py, most likely due to the fact that the magnetic energy of the nanopillar is minimized by reducing the magnetic flux due to stray fields\cite{45}. The total magnetic energy of the nanopillar will therefore be minimised by trying to align the Co and Py antiparallel to reduce magnetostatic fields. This behavior is confirmed from measurements of \textit{I}$_{\mathrm c}$(\textit{H}) which tend to show a slow fall-off of I$_{c}$ with H and the maximum I$_{c}$ is near H = 0 implying negligible barrier flux.

 These junctions follow the standard S/F'/F/F'/S device geometry proposed by Houzet and Buzdin\cite{20}, but in comparison to previous reports\cite{11,15,17} the source of non-collinearity in these junctions is the non-collinear magnetic structure formed in Py. 
 
\textbf{Exchange-spring junction characteristics.} 
Figure 3(a) shows \textit{I}$_{\mathrm c}$(\textit{H}) of a Nb/Co(2 nm)/Py(11 nm)/Nb asymmetric junction at 4.2 K, in which \textit{I}$_{\mathrm c}$(\textit{H})  is modulated confirming Josephson coupling. The maxima in \textit{I}$_{\mathrm c}$ occurs at \textit{H} = 0 to $\pm$2 mT due to intrinsic barrier flux from the barrier magnetization. The inset of Fig. 3(a) shows the low field regime of \textit{I}$_{\mathrm c}$(\textit{H}), emphasizing this shift. The shift in central peak in \textit{I}$_{\mathrm c}$ in each field direction is due to the changing magnetization of the composite barrier. The long range nature of supercurrents indicates that the transport  is via triplet pairs. 

Figure 3(b) shows \textit{R}($\theta$) for a Nb/Co(2 nm)/Py(11 nm)/Nb Josephson junction at 1.6 K using field values of 400 mT, 20 mT, 0 mT. R is defined as the junction resistance at a current bias slightly larger than \textit{I}$_{\mathrm c}$. Fig. 3(c), (d) and (e) represent the structure of magnetic moments corresponding to 400 mT, 20 mT and 0 mT. At 400 mT, the moments of Co and Py align in the direction of applied field. Therefore, R approaches the normal state of the junction and the \textit{I}$_{\mathrm c}$ = 0. At 20 mT, a non-collinear structure of magnetic moments is expected [Fig. 3(d)]. A modulation in R is observed with $\theta$; \textit{R}($\theta$) is a reflection of \textit{I}$_{\mathrm c}$($\theta$) in these junctions for bias currents just above the critical current. To illustrate this, we have shown \textit{I}$_{\mathrm c}$(\textit{H}) and $R(H)$ curves for a Nb/Co(2 nm)/Py(7 nm)/Nb
junction at 4.2 K in supplementary Fig. S1. The minima in \textit{I}$_{\mathrm c}$ corresponds to the maxima in resistance
and vice-versa. The \textit{R}($\theta$) results are consistent with the results obtained in the case of double XS junctions in Fig. 2(b). At 0 mT, the resistance is at a minimum and hence \textit{I}$_{\mathrm c}$ is a maximum. This can be explained due to a remnant non-collinear twist at 0 field [Fig. 3(e)], resulting in a higher critical current.

\section*{Discussion}
 We also investigated Nb/Py(6nm)/Nb control Josephson junctions in which no magnetic inhomogeneity is present. As shown in Fig. 4(a), there is no evidence for Josephson coupling in this junction, suggesting that the supercurrent in Nb/Co(2 nm)/Py(7 nm)/Nb is unconventional. The comparison of $R(H)$ curves of these two junctions (Supplementary Fig. S2) also agrees with the above observation. A further set of control junctions were fabricated in which Co and Py are magnetically decoupled by inserting a layer of 4-nm-thick Cu. Fig. 4(b) shows $I(V)$ curves for Nb/Co(2 nm)/Py(11 nm)/Nb and Nb/Co(2 nm)/Cu(4 nm)/Py(11 nm)/Nb Josephson junction. In the junctions with Cu (4 nm), no evidence for supercurrents is observed indicating that direct coupling between Py and Co is essential for establishing a long-range Josephson effect.

Figure 5 shows \textit{I}$_{\mathrm c}$\textit{R}$_{\mathrm N}$ vs total Py thickness for a series of junctions with magnetic barriers of Co/Py, Py/Co/Py, Py and Co/Cu/Py. $R_{\mathrm N}$ is measured at $I>>I_{\mathrm c}$ where the slope of the \textit{I}(\textit{V}) curve is constant. \textit{I}$_{\mathrm c}$\textit{R}$_{\mathrm N}$ for Nb/Py/Nb junction is taken from Ref. 32. We first discuss the case of Josephson junctions with Co/Py magnetic barriers. We observe that the decay of \textit{I}$_{\mathrm c}$\textit{R}$_{\mathrm N}$ with Py thickness is much slower compared to the Py-only junctions, indicating an additional component of supercurrent in XS junctions due to triplet pairs. Insertion of Cu between Co and Py magnetically isolates Py and Co and suppresses XS behavior and hence \textit{I}$_{\mathrm c}$ is zero.

  \textit{I}$_{\mathrm c}$\textit{R}$_{\mathrm N}$ for Co/Py junctions reduces to zero beyond 11 nm of Py, most likely due to the destruction of the spin-triplet correlations beyond the spin diffusion length ($\sim$ 5-10 nm)\cite{44, 41} of Py. The spin diffusion length is the distance over which electrons maintain spin-orientation. In the dirty limit, the decay length of spin-singlet supercurrents in F is given by\cite{1} $\xi_{F}=\sqrt{\hbar D_{F}/E_{ex}}$, where $D_{\mathrm F}$ is the diffusion coefficient and $E_{\mathrm ex}$ is the exchange energy within F. From previous reports\cite{10,11}, the decay length of spin-triplet supercurrents is limited by the spin diffusion length, $L_{F}=\sqrt{D_{F}\tau_{sf}}$ which is $\sim5-10$ nm in Py\cite{41,44} and $\sim$60 nm in Co\cite{44}. 
 
  We now compare \textit{I}$_{\mathrm c}$\textit{R}$_{\mathrm N}$ of asymmetric Nb/Co/Py/Nb with symmetric Nb/Py/Co/Py/Nb junctions. In Nb/Py/Co/Py/Nb junctions, there is a magnetically non-collinear structure at both Nb/Py interfaces. We observe a slow decay of \textit{I}$_{\mathrm c}$\textit{R}$_{\mathrm N}$ for symmetric as well as for asymmetric junctions. A finite supercurrent is observed up to 16 nm of total ferromagnetic thickness [Py(7 nm)/Co(2 nm)/Py(7 nm)] in symmetric junctions as shown in the inset of Fig. 5. This shows that the supercurrent observed in Josephson junctions with symmetric as well as asymmetric Co/Py XS interfaces cannot be solely singlet supercurrent and must originate from triplet Cooper pairs.

\section*{Conclusion}
We have demonstrated triplet-supercurrent-control with magnetic field orientation in junctions containing Co/Py XS interfaces. XS interfaces pave the way to future experiments where the ground-state phase difference across the junction can be controlled by changing magnetic configuration. Another key finding of this work is that the triplet supercurrents exists in asymmetric as well as symmetric Josephson junctions. Finally, supercurrents are detected in Py over $\sim$11 nm, which is comparable to the spin-diffusion length in Py of (5-10 nm)\cite{41,44}.

\begin{methods}
\textbf{Film Growth.} Thin films of Nb(220 nm)/Co(2 nm)/Py(0-13 nm)/Nb(220 nm) and Nb(220 nm)/Py(1-13)/Co(2 nm)/Py(1-13)/Nb(220 nm) were prepared at room temperature in an Ar pressure of 1.5 Pa, by dc-magnetron sputtering using high purity ($99.95\%$) Nb, Co and Py (Ni$_{80}$Fe$_{20}$) targets on Si/SiO$_2$ substrates. The base pressure of the deposition system was of the order of $10^{-9}$ mBar. Film thicknesses were calibrated by growing films onto patterned substrates and by measuring the height of step edges with an atomic force microscope. Film thicknesses were confirmed using low angle X-ray reflectivity. The entire series of films was prepared in a single deposition run, ensuring uniform interface quality. 

\textbf{Device Fabrication.} Nanopillar Josephson junctions were fabricated by defining 4-$\mu$m-wide and 30-$\mu$m-long tracks using optical lithography and Ar-ion-milling, and Ga-ion Focused Ion Beam (FIB) milling process was used to define the junction\cite{43}. Junction dimensions varied from (300$\times$300) nm$^{2}$ to (300$\times$500) nm$^{2}$. 

\textbf{Transport measurements.} Magnetization (\textit{M}) measurements were made using a Quantum Design Squid magnetometer with magnetic fields ($\mu_{0}H$) applied in-plane. Junction measurements were performed in a $\mu$-metal shielded dipstick probe in a liquid He dewar and cryogen-free pump probe system. Current-voltage \textit{I}(\textit{V}) characteristic were measured in a four probe current-biased configuration.

\end{methods}


\begin{addendum}
	\item We thank Dr. Muhammad Shahbaz Anwar for useful discussions. J.W.A. Robinson acknowledges funding from the Royal Society and the EPSRC through a Programme Grant (EP/M50807/1) and International Network (EP/P026311/1). K. Senapati acknowledges funding from National Institute of Science Education and Research (NISER), DST-Nanomission (SR/NM/NS-1183/2013) and DST-SERB (EMR/2016/005518) of Govt. of India. 
	\item[Competing Interests] The authors declare that they have no
	competing financial interests.
	\item[Author contribution] EB, KS and JWAR planned the experiment. EB performed most of the device fabrication and measurements. JDS and SK helped for device fabrication and measurements. JDS, AS and NAS helped during sputtering of multilayer films. ZHB, KS and JWAR supervised the work. All authors contributed to the manuscript preparation and corrections.   
	\item[Correspondence] Correspondence and requests for materials
	should be addressed to J. W. A. Robinson.
	(email: jjr33@cam.ac.uk) and K.Senapati. (email: kartik@niser.ac.in).
\end{addendum}

\newpage 

\begin{figure}[htbp!]
\centering
	\includegraphics[width=8cm]{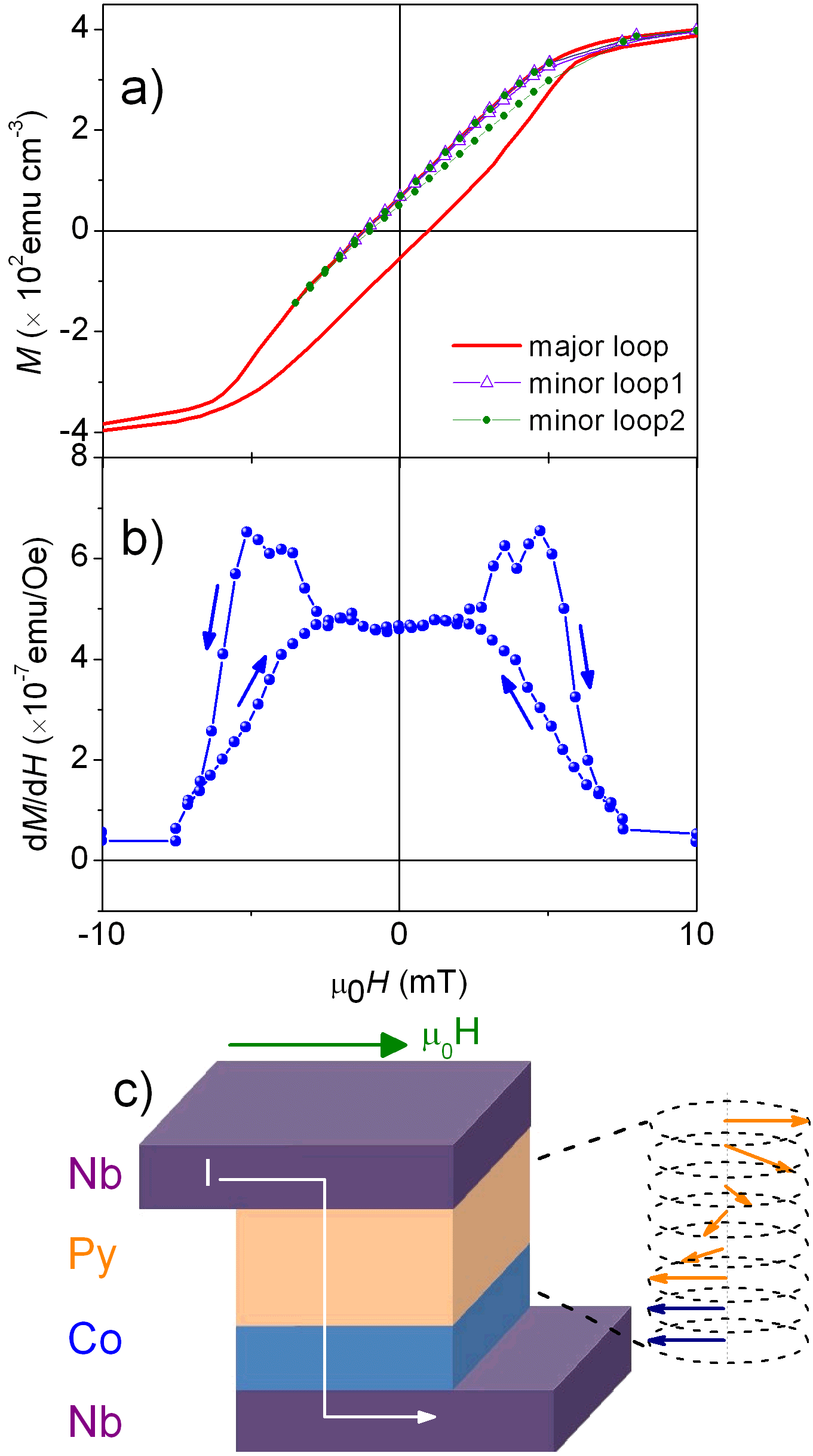}\\
	\caption{\textbf{Magnetic characterization of a Co/Py XS}. (a) Major and minor \textit{M}(\textit{H}) loops of an unpatterned Nb(220 nm)/Co(2 nm)/Py(11 nm)/Nb(220 nm) multilayer at 10 K with H in-plane. For a range of magnetic fields the minor loops are reversible, consistent with exchange-spring behavior. (b) The field derivative curve of an unpatterned Nb/Co(2 nm)/Py(11 nm)/Nb multilayer at 10 K. (c) Schematic illustration of a nanopillar Josephson junction with an XS barrier showing the expected nanomagnetic structure. }
\end{figure}

\newpage	
\begin{figure}[htbp!]
		\includegraphics[width=15cm]{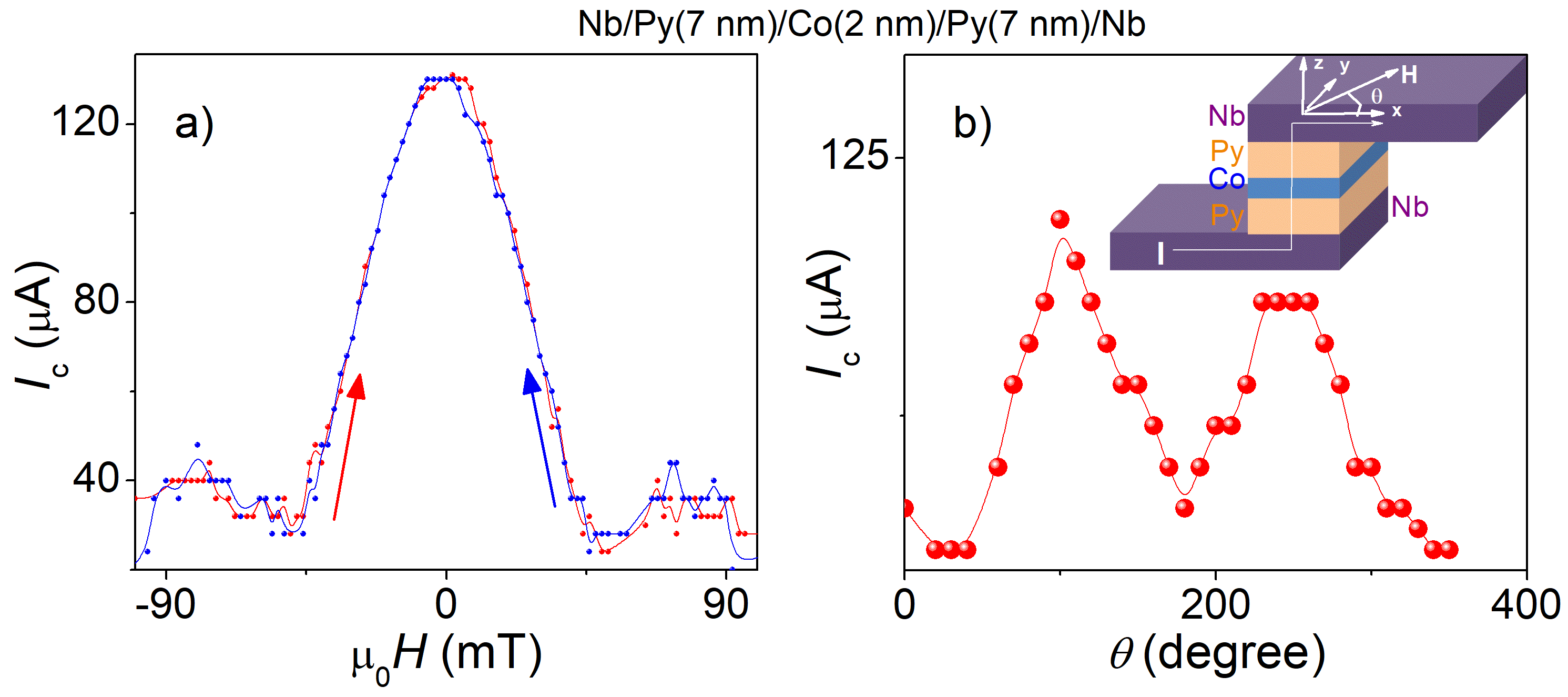}\\
		\caption{\textbf{Characterization of a Nb/Py(7 nm)/Co(2 nm)/Py(7 nm)/Nb double XS Josephson junction}. (a) Fraunhofer modulation of $I_{\mathrm c}(H)$ of double XS junction at 1.6 K. (b) Critical current vs direction of applied magnetic field showing manipulation of triplet supercurrents for double XS junction with dimensions of $\sim$(300$\times$300) nm$^2$ at 1.6 K with in-plane field 20 mT, at different angles ($\theta$) of the applied field with respect to the length of the track (inset: \textit{x}-axis). Inset is a schematic illustration of a junction showing the direction of bias currents \textit{I} and applied magnetic field \textit{H}.}
\end{figure}

\newpage
\begin{figure}[htbp!]
\centering
		\includegraphics[width=15cm]{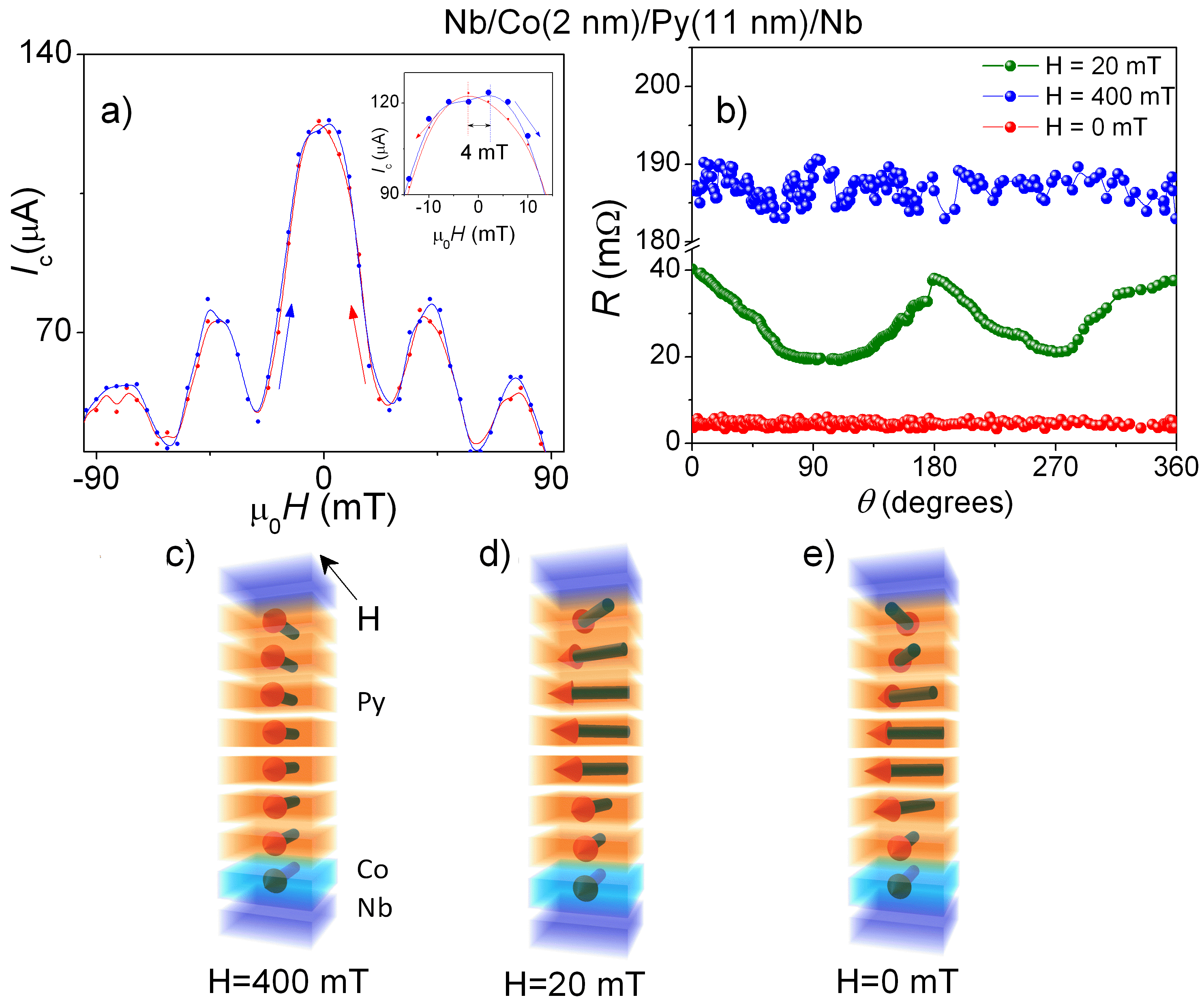}\\
		\caption{\textbf{Characterization of Nb/Co(2 nm)/Py(11 nm)/Nb XS Josephson junction} (a) Fraunhofer modulation of $I_{\mathrm c}(H)$ of a XS junction at 4.2 K. Inset shows the low field regime of $I_{\mathrm c}(H)$, with hysteresis. (b) R vs direction of applied magnetic field showing manipulation of triplet supercurrents in the same sample with dimensions of $\sim$(300$\times$300) nm$^2$ at 1.6 K with in-plane fields of 400 mT, 20 mT and 0 mT at different angles ($\theta$) of the applied field with respect to the length of the track. (c) Schematic illustration showing the moments of Co and Py aligned at 400 mT. (d) At 20 mT, a non-collinear magnetic structure forms in Co/Py. (e) At 0 mT, a residual magnetic non-collinearity is present. }
\end{figure}

\newpage
\begin{figure}[htbp!]
\centering
	\includegraphics[width=18cm]{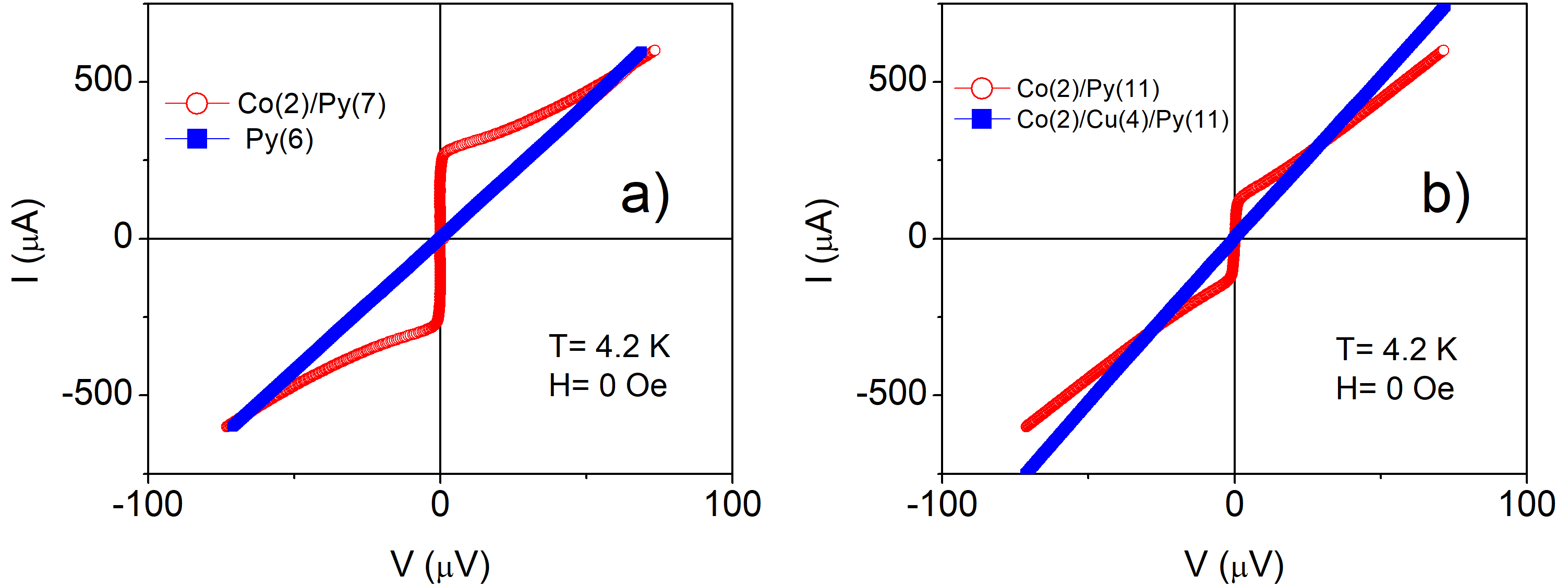}\\
	\caption{\textbf{Comparisons showing the long range nature of supercurrent in a S/XS/S magnetic Josephson junction} (a) \textit{I(V)} characteristics of a Nb(220 nm)/Co(2 nm)/Py(7 nm)/Nb(220 nm) junction and a Nb(220 nm)/Py(6 nm)/Nb(220 nm) junction at 4.2 K and zero magnetic field. (b) Comparison of I(V) of a Nb/Co(2 nm)/Py(11 nm)/Nb junction with a Nb(220 nm)/Co(2 nm)/Cu(4 nm)/Py(11 nm)/Nb(220 nm) junction at 4.2 K and zero field.}
\end{figure}

\newpage			
\begin{figure}[htbp!]
\centering
	\includegraphics[width=10cm]{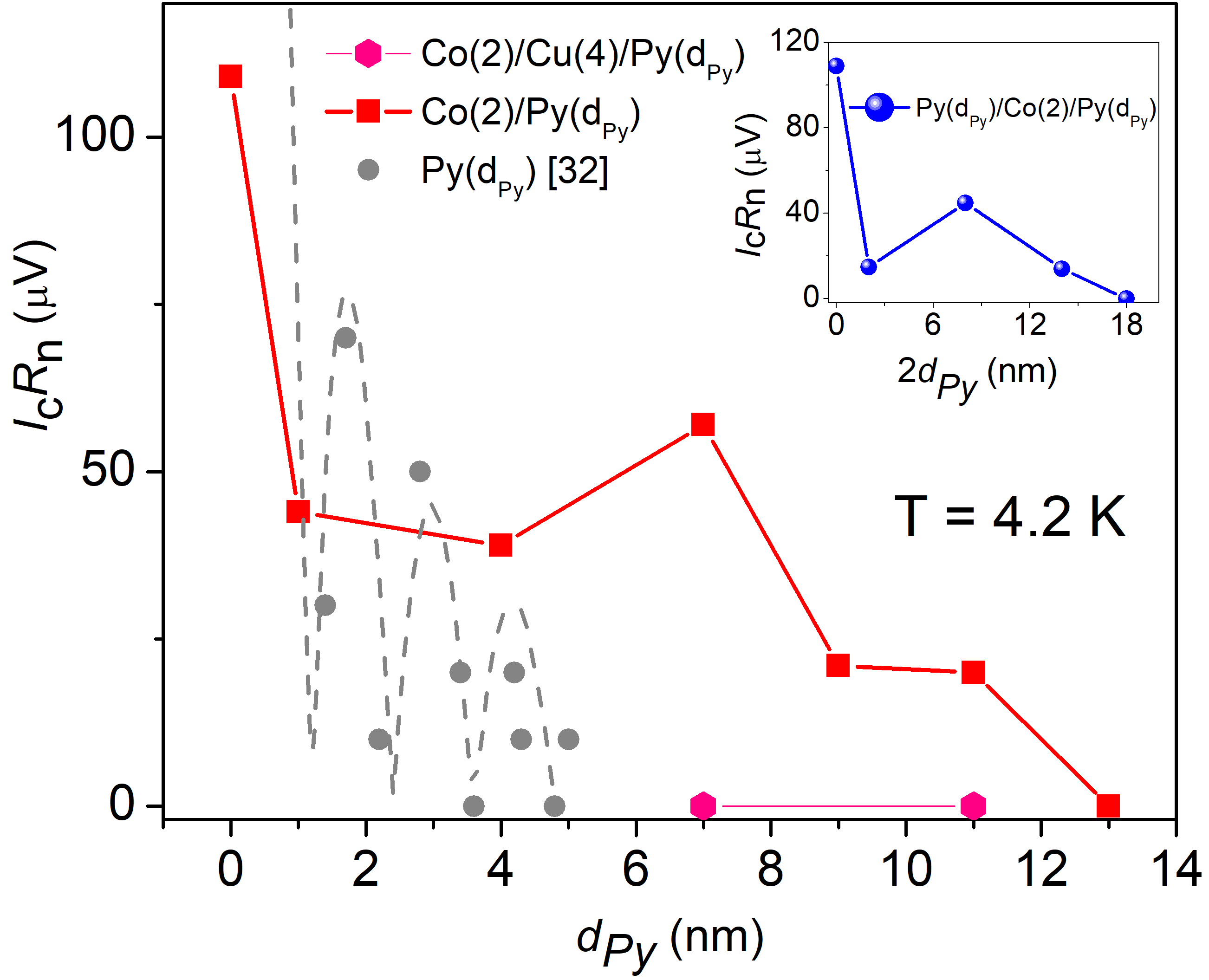}\\
	\caption{\textbf{Supercurrent decays in Py}. $I_{\mathrm c}R_{\mathrm N}$ vs. total thickness of Py; the grey curve presents the $I_{\mathrm c}R_{\mathrm N}$ for Py-only junctions (taken from Ref. 32). As $d_{\mathrm Py}$ increases above 5 nm, $I_{\mathrm c}R_{\mathrm N}$ is practically zero but introducing Co results in an enhancement of $I_{\mathrm c}R_{\mathrm N}$ which decays slowly. Inset shows $I_{\mathrm c}R_{\mathrm N}$ for a double spring junction vs total thickness of Py.}
\end{figure}
\newpage
\beginsupplement
\title{\centering Supplementary figures\\ }
\maketitle
\vskip1cm

\begin{figure}[htbp!]
	\includegraphics[width=15 cm]{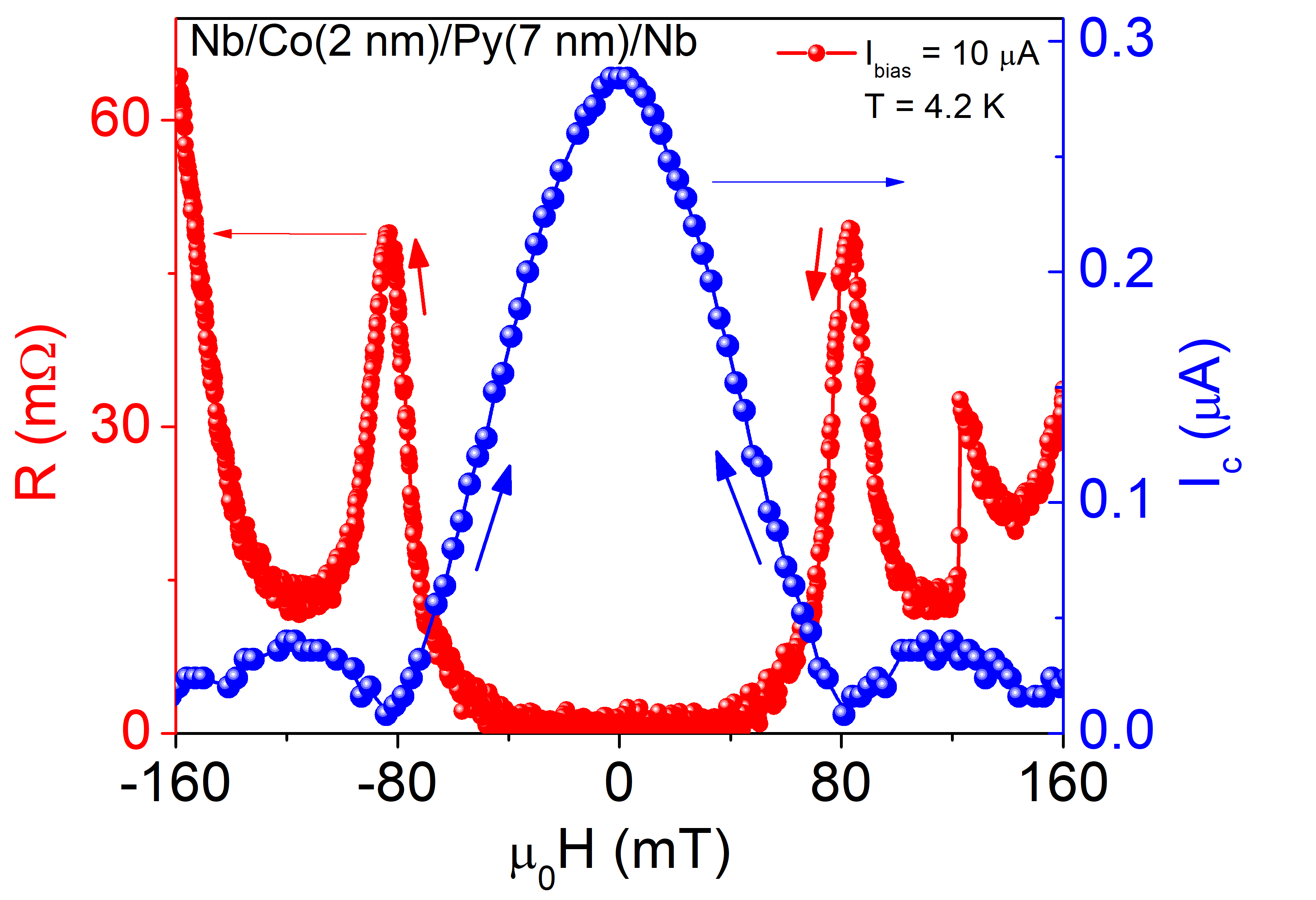}
	\caption{\textbf{Comparison of I$_ c$(H) and R(H) for Nb/Co(2 nm)/Py(7 nm)/Nb Josephson junction}. Red points on left hand axis represent the R(H) curve while blue points on right axis represent the I$_c$(H). The minima in I$_c$(H) and the maxima in R(H) lie at same magnetic field, signifying the equivalence of both measurements.}
\end{figure}
\ref{fig:picture}

\begin{figure}[htbp!]
	
	\includegraphics[width=12 cm]{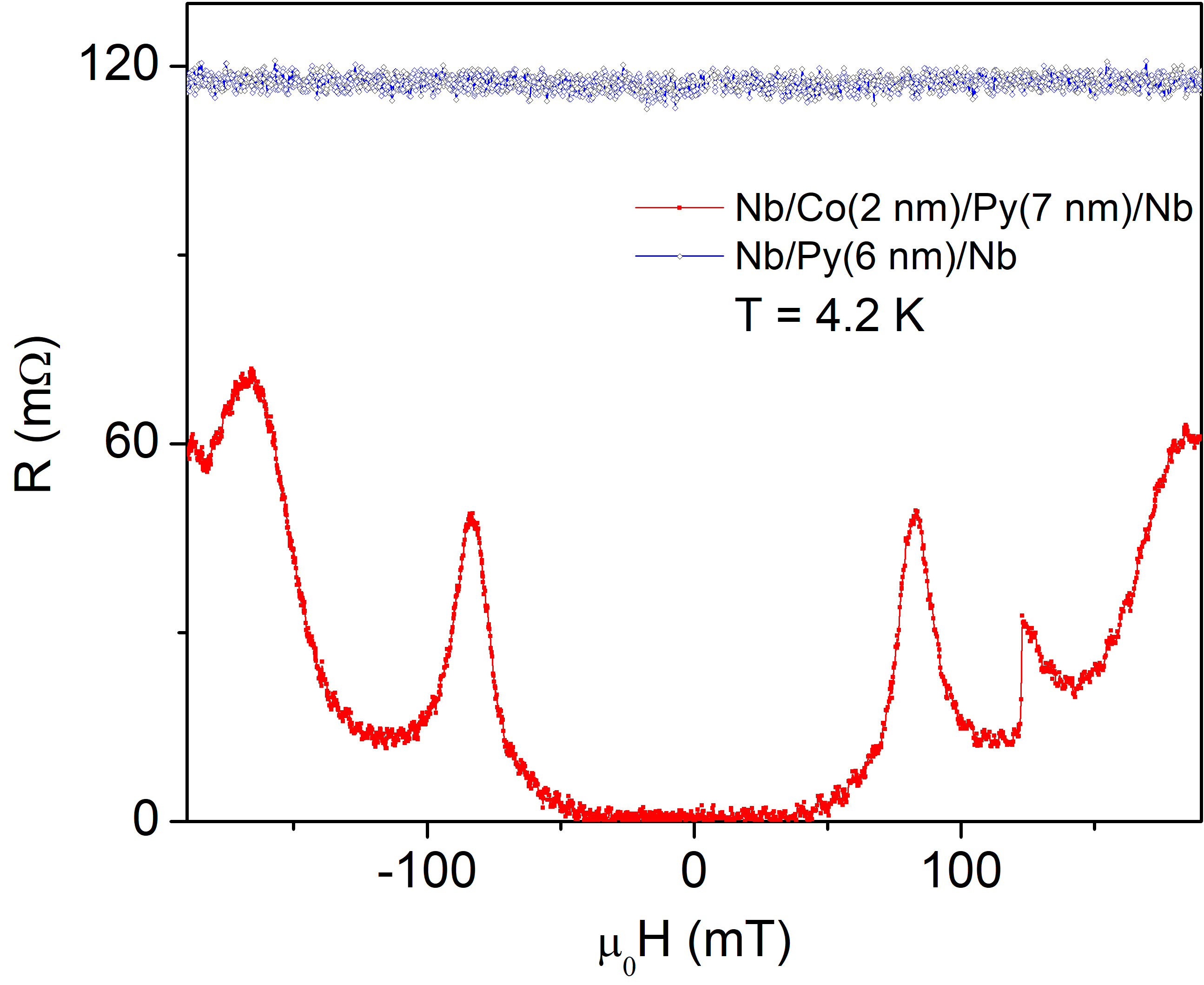}
	\caption{\textbf{Comparison of R(H) curves for Nb/Co(2 nm)/Py(7 nm)/Nb and
			Nb/Py(6 nm)/Nb Josephson junctions at 4.2 K}. Fraunhofer type modulations were observed in junction resistance as a function of magnetic field for the Nb/Co(2 nm)/Py(7 nm)/Nb while no field dependence was observed for Nb/Py(6 nm)/Nb junction.}
\end{figure}

\end{document}